\documentclass[twocolumn,english,aps,prd,graphics,floatfix,a4paper,superscriptaddress]{revtex4-2} 
\usepackage{amsmath, amssymb}
\usepackage{makecell}
\usepackage{multirow}
\usepackage{CJK}
\usepackage{graphicx}
\usepackage{mathrsfs}
\usepackage{bm}
\usepackage{amsmath}
\usepackage{dcolumn}
\usepackage{epstopdf}
\usepackage{dsfont}
\usepackage{amssymb}
\usepackage{tabularx}
\usepackage{array}
\usepackage{float}
\usepackage{color}
\usepackage{epstopdf}
\usepackage{mathrsfs}
\usepackage[colorlinks, linkcolor=blue,anchorcolor=blue,citecolor=blue,urlcolor=blue]{hyperref}
\usepackage{extarrows}

\begin{document}

\title{Two-dimensional Janus Si dichalcogenides: A first-principles study}

\author{San-Dong Guo}
\email{sandongyuwang@163.com}
\thanks{These authors contributed equally to this work.}
\affiliation{School of Electronic Engineering, Xi'an University of Posts and Telecommunications, Xi'an 710121, China}

\author{Xu-Kun Feng}
\thanks{These authors contributed equally to this work.}
\affiliation{Research Laboratory for Quantum Materials, Singapore University of Technology and Design, Singapore 487372, Singapore}

\author{Yu-Tong Zhu}
\affiliation{School of Electronic Engineering, Xi'an University of Posts and Telecommunications, Xi'an 710121, China}

\author{Guangzhao Wang}
\affiliation{Key Laboratory of Extraordinary Bond Engineering and Advanced Materials Technology of Chongqing, School of Electronic Information Engineering, Yangtze Normal University, Chongqing 408100, China}

\author{Shengyuan A. Yang}
\affiliation{Research Laboratory for Quantum Materials, Singapore University of Technology and Design, Singapore 487372, Singapore}

\begin{abstract}
Strong structural asymmetry is actively explored in two-dimensional (2D) materials, because it can give rise to many interesting physical properties. Motivated by the recent synthesis of monolayer $\mathrm{Si_2Te_2}$, we explore a family of 2D materials, termed as the Janus Si dichalcogenides (JSD), which parallel the Janus transition metal dichalcogenides and exhibit even stronger inversion asymmetry. Using first-principles calculations, we demonstrate excellent stability of these materials. We show that their strong structural asymmetry leads to pronounced intrinsic polar field, sizable spin splitting due to spin-orbit coupling, and large piezoelectric response. The spin splitting involves an out-of-plane component, which is beyond the linear Rashba model. The piezoelectric tensor has large value in both in-plane $d_{11}$ coefficient and out-of-plane $d_{31}$ coefficient, making the monolayer JSDs distinct among the existing 2D piezoelectrics. In addition, we find interesting strain-induced phase transitions in these materials. Particularly, there are multiple valleys in the conduction band that compete for the conduction band minimum, which will lead to notable changes in optical and transport properties under strain. Our work reveals a new family of Si based 2D materials, which could find promising applications in spintronic and piezoelectric devices.

\end{abstract}

\maketitle

\section{Introduction}

The field of two-dimensional (2D) materials has been rapidly expanding in recent years, driven by the continued realization of new materials and discovery of their novel properties~\cite{ys1,ys2,ys3,ys4,ys5,q4,q4-1,q4-2}. A material's property is closely connected to its symmetry. While a lot of 2D materials tend to crystallize in high-symmetry structures, there also exist crystals with intrinsic structural asymmetry or exhibiting spontaneous symmetry breaking. Such 2D materials with lower symmetry can be even more interesting, because they can host physical effects that are forbidden in high-symmetry structures. For example, piezoelectricity, which represents the coupling between electric polarization and strain/stress and is important for a wide range of applications such as sensors, actuators, and energy harvesters~\cite{q4,q4-1,q4-2}, necessarily requires the underlying crystals to have broken inversion symmetry. It is also well known that structural inversion asymmetry can help to enhance the Rashba-type spin-orbit coupling (SOC) in 2D systems~\cite{y1}, which is desired for spintronics applications. These points clearly manifest in the Janus monolayer transition metal dichalcogenides (TMDs)~\cite{ys6,ys7,ys8,y2,y3,y4}. In these materials, the symmetry is lowered by replacing one layer of chalcogen atoms by another species in the same group. For example, starting from monolayer MoSe$_2$ (MoS$_2$) on a substrate, one can achieve
monolayer MoSSe via controlled sulfurization (selenization) of the top atomic layer, as successfully demonstrated in experiment~\cite{e2,e1}. Previous studies have indeed shown that monolayer MoSSe exhibits Rashba-type SOC and has large piezoelectric response, especially a nonzero out-of-plane piezoelectric coefficient $d_{31}\sim 0.02$ pm/V~\cite{q5-11}.

To incorporate 2D materials into the existing semiconductor technology, Si-based materials are most desired. In a very recent experiment, a Si-based 2D material, the monolayer Si$_2$Te$_2$, was successfully synthesized on a substrate and shown to be a semiconductor~\cite{y8}. The crystal preserves inversion symmetry, so it does not allow any intrinsic piezoelectricity nor Rashba SOC. Nevertheless, one notes that the structure contains the chalcogen Te atoms forming the two outer layers, similar to the TMDs. Therefore, it is possible to use similar techniques as for synthesizing Janus TMDs~\cite{e2,e1} and to convert  monolayer Si$_2$Te$_2$ into a Janus Si dichalcogenides (JSD) monolayer.

In this work, we explore this idea and perform a systematic study on the  monolayer JSDs, using first-principles calculations. We show that these materials, namely Si$_2$STe, Si$_2$SeTe, and Si$_2$SSe, are all stable in the Janus monolayer structure. They enjoy good thermal and mechanical stability. Using monolayer Si$_2$SeTe as an example, we show that its internal effective electric field along the vertical direction can reach $\sim 1.8$ V/\AA, which is about twice of monolayer MoSSe. Importantly, the formation of Janus structure breaks the inversion symmetry of monolayer Si$_2$Te$_2$, leading to a sizable SOC splitting in the band structure  and a large piezoelectric effect. For the SOC splitting, we show that the split bands also possess a sizable spin-$z$ components, which requires higher-order terms beyond the simple linear Rashba model. As for piezoelectricity, we find that the in-plane coefficient $d_{11}$ in Si$_2$SeTe is larger than most existing 2D materials, and it also allows a sizable out-of-plane coefficient $d_{31}$ which is an order of magnitude larger than Janus TMDs. These values are even more enhanced in Si$_2$STe due to the larger contrast between the two chalcogen species. In addition, we show that a moderate strain on monolayer JSDs can induce a significant change of the band structure, leading to phase transitions and changes in the conduction band minimum (CBM). Our work reveals a new family of 2D semiconductors, which have great potential for spintronic and piezoelectric device applications.

\begin{table*}
\centering \caption{Structural and elastic parameters of Janus monolayer Si$_2$XY (X, Y=S, Se, Te) materials. These include lattice constant $a_0$ ($\mathrm{{\AA}}$),  Si-X ($d_1$) and Si-Y ($d_2$) bond lengths ($\mathrm{{\AA}}$),
X-Si-X ($\theta_1$) and Y-Si-Y ($\theta_2$) bond angles ($^{\circ}$),  layer thickness ($t$) ($\mathrm{{\AA}}$),
elastic constant $C_{ij}$ ($\mathrm{Nm^{-1}}$), shear modulus
$G_\text{2D}$ ($\mathrm{Nm^{-1}}$),  Young's modulus $C_\text{2D}$  ($\mathrm{Nm^{-1}}$), and  Poisson's ratio $\nu_\text{2D}$. }\label{tab0}
  \begin{tabular*}{0.96\textwidth}{@{\extracolsep{\fill}}lccccccccccc}
  \hline\hline
Name&$a_0$& $d_1$ & $d_2$& $\theta_1$&$\theta_2$&$t$&$C_{11}$/$C_{22}$ &  $C_{12}$& $G_\text{2D}$&$C_\text{2D}$& $\nu_\text{2D}$\\\hline
$\mathrm{Si_2SSe}$&3.537& 2.35 & 2.45 & 97.55 & 92.43 &4.18&92.67&32.88&29.90&81.00&0.36  \\
$\mathrm{Si_2STe}$&3.684& 2.40 & 2.64 & 100.35 & 88.64 &4.28&84.59&36.00&24.30&69.27&0.43  \\
$\mathrm{Si_2SeTe}$&3.763& 2.52 & 2.66 & 96.63 & 90.13 &4.38&75.69&33.96&20.87&60.45&0.45  \\\hline\hline
\end{tabular*}
\end{table*}

\section{Computation method}
Our first-principles calculations were based on the density functional theory (DFT)~\cite{1}, using  the projector augmented wave method  as implemented in
the  Vienna ab initio Simulation Package (VASP)~\cite{pv1,pv2,pv3}.  The  generalized gradient approximation with Perdew, Burke and  Ernzerhof (PBE) realization was adopted to treat the exchange-correlation potential~\cite{pbe}. The kinetic  energy cutoff was set to 500 eV.
The energy and force convergence criteria were set to $10^{-8}$ eV and $10^{-4}$ $\mathrm{eV\cdot {\AA}^{-1}}$. A vacuum spacing larger than $17$ $\mathrm{{\AA}}$ along the $z$ direction is included to suppress artificial interactions
between periodic images.

The phonon spectra were calculated using the Phonopy code~\cite{pv5}, with a supercell
of $5\times5\times1$.
The ab initio molecular dynamics (AIMD) simulations were performed  by using the canonical ensemble with a $4\times4\times 1$ supercell. The simulated duration was set to be 8 ps
with a time step of 1 fs.
The elastic stiffness tensor $C_{\ell k}$ and  piezoelectric stress tensor $e_{i\ell}$ were calculated by using the strain-stress relationship and density functional perturbation theory (DFPT) method~\cite{pv6}.
Note that to obtain the tensor elements for a 2D system, a renormalization by the $z$ lattice parameter (i.e., the spacing two neighboring layers) is needed in the calculation~\cite{q5,q5-1}.
A $\Gamma$-centered $18\times18\times1$ $k$-point mesh  was adopted in the calculation of  $C_{\ell k}$ as well as the electronic structures, and a $10\times18\times1$ Monkhorst-Pack $k$-point mesh was used for
$e_{i\ell}$ (due to the change of the unit cell).
The PYPROCAR code was used to obtain the constant energy contour plots of the spin
polarization~\cite{py}.

\begin{figure}
  \includegraphics[width=8.4cm]{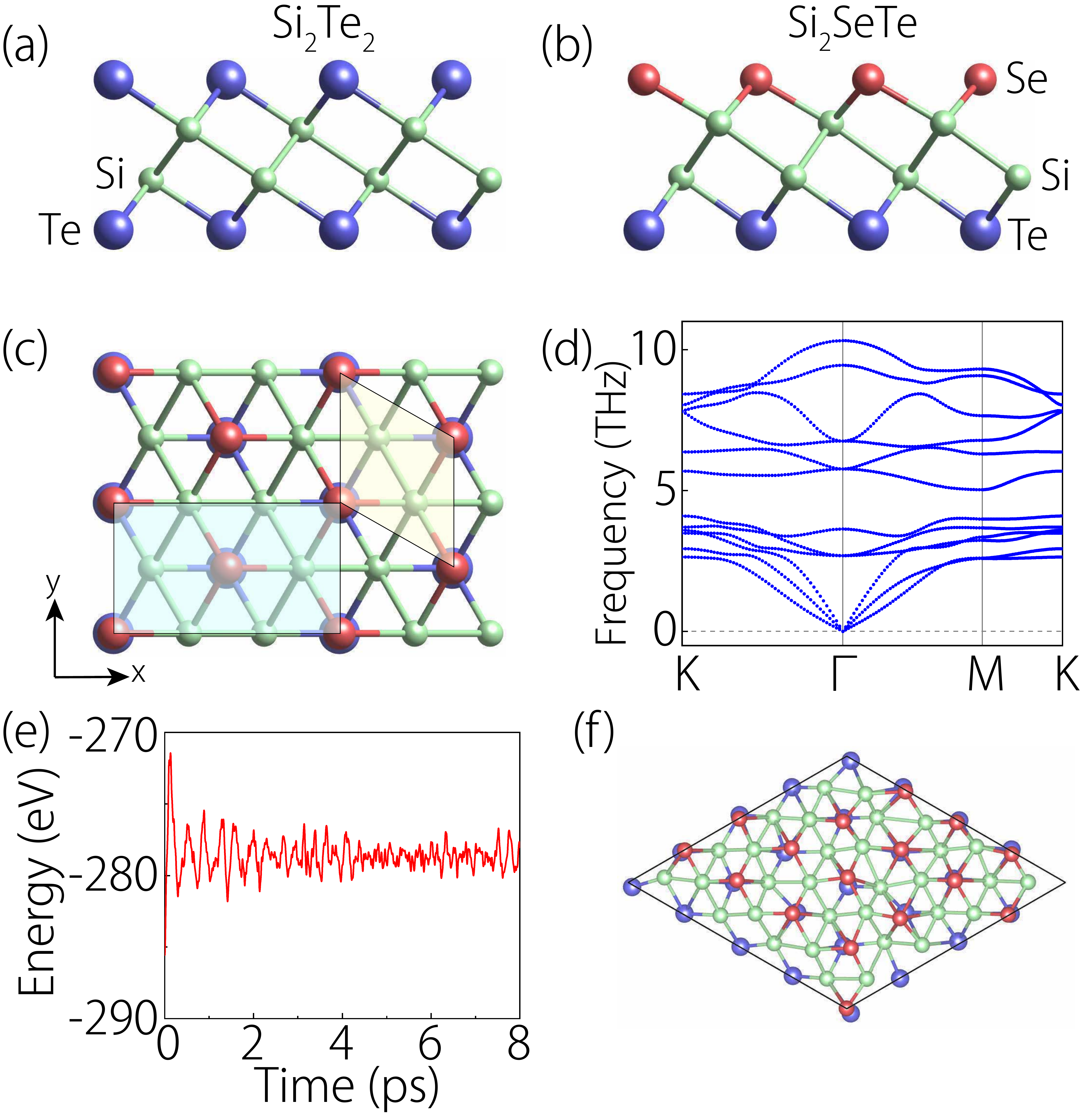}
  \caption{Crystal structure (side view) of (a) monolayer Si$_2$Te$_2$ and (b) Janus monolayer  $\mathrm{Si_2SeTe}$. (c) shows the top view of $\mathrm{Si_2SeTe}$.  The yellow shaded region shows the primitive cell, while the green shaded region shows the rectangular unit cell. (d) Calculated phonon spectrum of  $\mathrm{Si_2SeTe}$. (e) and (f) show the AIMD result on $\mathrm{Si_2SeTe}$ at 800 K. (e) Energy fluctuation during the AIMD simulation. (f) is the snapshot of the lattice at the end of the simulation.}\label{fig1}
\end{figure}

\section{Structure and stability}

As illustrated in Fig.~\ref{fig1}(a), the experimentally synthesized monolayer Si$_2$Te$_2$ has a hexagonal lattice structure with a space group of $P\bar{3}m1$ (No.~164)~\cite{y8}. It consists of four atomic layers stacked in the sequence of Te-Si-Si-Te. The JSD monolayer is obtained by replacing the top Te layer by a another chalcogen element, e.g., S or Se, which leads to monolayer Si$_2$STe and Si$_2$SeTe, as shown in Fig.~\ref{fig1}(b,c). For completeness, we also consider the closely related JSD material Si$_2$SSe here. Their optimized lattice parameters are listed in Table~I. One can see that these parameters are slightly smaller compared to monolayer Si$_2$Te$_2$ ($\sim 3.99$ \AA)~\cite{y9,y10}, which can be understood because S and Se have smaller radius than Te and their electronegativity is also larger. Because the three JSDs share similar behavior, in the following, we shall mainly focus on Si$_2$SeTe. The results for the other two are briefly mentioned or put in the Supplemental Material~\cite{bc}.

The formation of Janus structure lowers the symmetry of the crystal to $P3m1$ (No.~156). The most important change is that the inversion symmetry $\mathcal{P}$ originally preserved in monolayer Si$_2$Te$_2$ is broken in monolayer JSDs, and the system changes from a non-polar crystal to a polar crystal. This underlies the enhanced spin splitting and piezoelectricity to be discussed in a while.

To check the stability of these 2D materials, we compute their phonon spectra. Figure~\ref{fig1}(d) plots the result for monolayer
Si$_2$SeTe. One can see there is no imaginary frequency modes, which indicates that the structure is dynamically stable.
In the spectrum, both
linear and flexural modes  can be observed  around the $\Gamma$ point, which shares the general feature of 2D materials~\cite{r1,zhu2014,r2}.

To investigate the thermal stability, we conduct the AIMD simulations with temperatures up to 1000 K. Figure~\ref{fig1}(e,f) shows the simulation result on monolayer Si$_2$SeTe at 800 K. One observes that the overall structure is well maintained at the end of the simulation time, which indicates its good thermal stability.

We further evaluate the elastic constants of monolayer Si$_2$SeTe. Here, the $x$, $y$ and $z$ axis are chosen to be along the armchair, zigzag, and vertical directions, respectively (see Fig.~\ref{fig1}(c)).  Using Voigt notation, the tensor takes the form of $C_{\ell k}$ with $\ell, k=1,\cdots,6$. However, for 2D materials, one usually considers stresses and strains only within the basal plane, whereas the $z$-direction is stress/strain free. This eliminates the elements with  $\ell$ or $k$ in $\{3,4,5\}$.
Further constrained by the $C_{3v}$ point group of monolayer JSD, the elastic tensor can be expressed in the form of~\cite{q5,q5-1,q5-11}
\begin{equation}\label{pe1-4}
   C=\left(
    \begin{array}{ccc}
      C_{11} & C_{12} & \cdot \\
     C_{12} & C_{11} & \cdot \\
      \cdot & \cdot & (C_{11}-C_{12})/2 \\
    \end{array}
  \right),
\end{equation}
where $\cdot$ denotes the zero element, and we omit the $3\times 3$ vanishing block corresponding to $\ell,k\in\{3,4,5\}$.
For monolayer Si$_2$SeTe, we find the calculated  $C_{11}=75.69$ $\mathrm{Nm^{-1}}$ and $C_{12}=33.96$ $\mathrm{Nm^{-1}}$, which satisfy the  Born  criteria~\cite{ela}:
$C_{11}>0$ and  $C_{11}-C_{12}>0$, confirming its mechanical stability.
The shear modulus $C_{66}=(C_{11}-C_{12})/2$ is 20.87 $\mathrm{Nm^{-1}}$.
The direction-dependent Young's modulus $C_\text{2D}(\theta)$ can be obtained as~\cite{ela1}
\begin{equation}\label{c2d}
C_\text{2D}(\theta)=\frac{C_{11}C_{22}-C_{12}^2}{C_{11}\sin^4\theta+A\sin^2\theta \cos^2\theta+C_{22}\cos^4\theta},
\end{equation}
where $\theta$ is the polar angle measured from $x$, and $A=(C_{11}C_{22}-C_{12}^2)/C_{66}-2C_{12}$. For $C_{3v}$ systems satisfying relation (\ref{pe1-4}), one easily finds that $C_\text{2D}$ is isotropic, as it should be, and
\begin{equation}
  C_\text{2D}=C_{11}-C_{12}^2/C_{11}.
\end{equation}
For monolayer  Si$_2$SeTe, the obtained $C_\text{2D}$ is 60.45 $\mathrm{Nm^{-1}}$.
This value is much smaller than graphene ($\sim 340\pm 40$ Nm$^{-1}$) and MoS$_2$ ($\sim 126.2$ Nm$^{-1}$)~\cite{q5-1,q5-1-1}, indicating the better mechanical flexibility of monolayer  Si$_2$SeTe. The Poisson's ratio $\nu_\text{2D}$ can also be derived from the elastic constants as
\begin{equation}\label{e1}
\nu_\text{2D}=\frac{C_{12}}{C_{11}},
\end{equation}
which is about 0.45 for monolayer  Si$_2$SeTe.

These material properties and the results for the other two JSD monolayers are listed in Table~I.

\section{Electronic structure and spin splitting}
\begin{figure}
  \includegraphics[width=7cm]{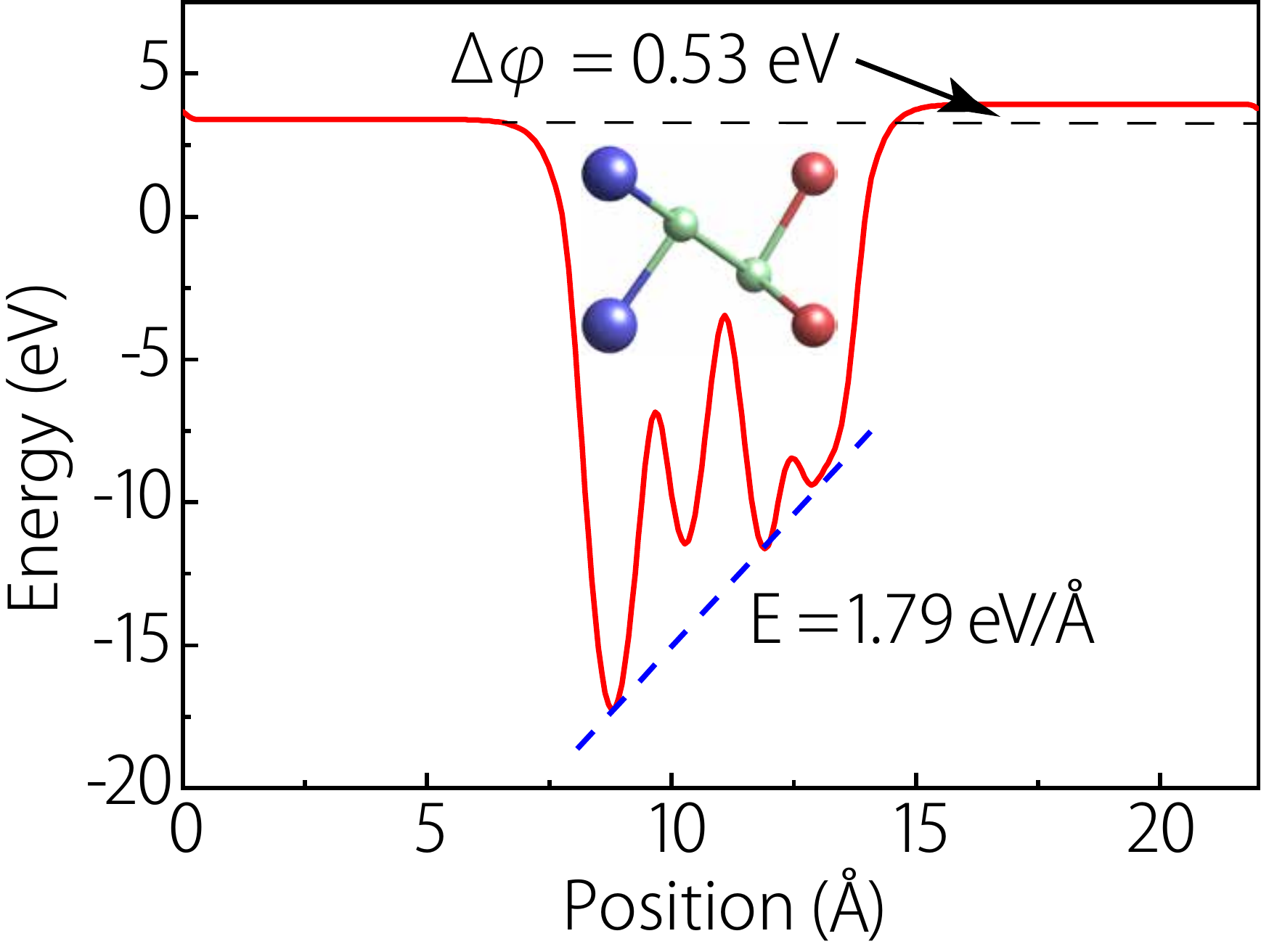}
\caption{Planar averaged electrostatic potential energy variation along $z$ for monolayer $\mathrm{Si_2SeTe}$. $\Delta\varphi$ is the potential energy difference across the layer. $E$ stands for the intrinsic polar field.}\label{fig2}
\end{figure}

Like Janus TMDs, monolayer JSDs possess an intrinsic polar electric field along the $z$ direction, due to the broken $\mathcal{P}$ symmetry and the different electronegativity of the two chalcogen elements on the two sides.
To better understand this field, in Fig.~\ref{fig2}, we plot the planar  average of the
electrostatic potential energy as a function of the $z$ coordinate for monolayer Si$_2$SeTe. One clearly observes an asymmetric distribution associated with the $\mathcal{P}$ breaking. This induces an electrostatic potential energy difference $\sim 0.53$ eV between the sides, which can be reflected as a surface dependent work function.  As shown in Fig.~\ref{fig2}, from the slope of the curve, we extract the strength of the intrinsic polar field to be about 1.79 $\mathrm{V/{\AA}}$. One notes that this value is more than two times that of MoSSe ($\sim 0.856$ $\mathrm{eV/{\AA}}$)~\cite{y11} and is much larger than the practical value that can be achieved by a gate field.  The stronger polar field implies more pronounced effects associated with
$\mathcal{P}$ symmetry breaking.

\begin{figure}
  \includegraphics[width=8cm]{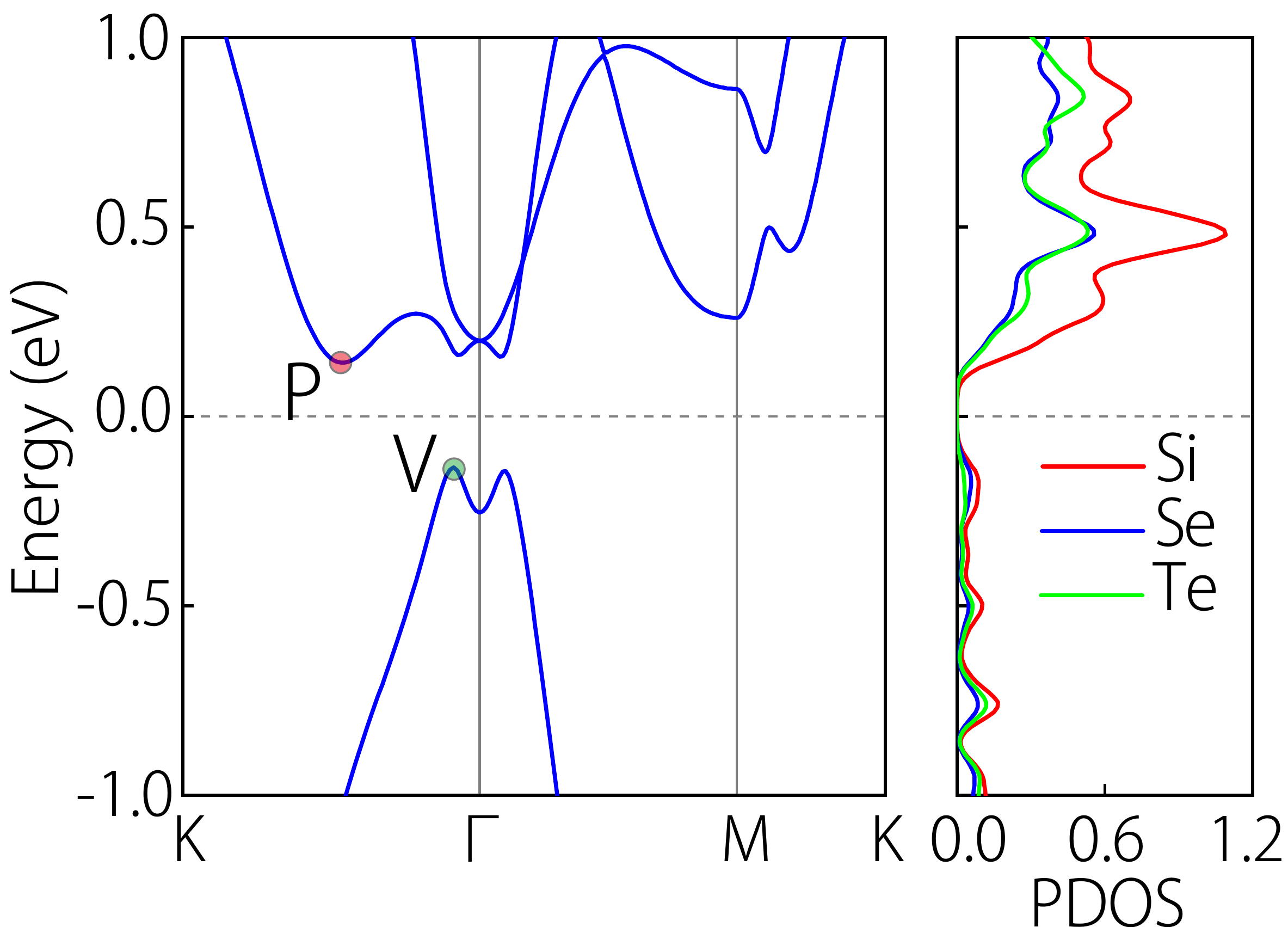}
\caption{Band structure and PDOS of monolayer $\mathrm{Si_2SeTe}$ in the absence of SOC. }\label{fig3}
\end{figure}

Next, we investigate the electronic band structure of monolayer Si$_2$SeTe. Figure~\ref{fig3} plots the band structure in the absence of SOC, along with the projected density of states (PDOS). One observes an indirect gap semiconductor with a band gap of about 276 meV. The low-energy bands are dominated by the Si $p$ orbitals. In Fig.~\ref{fig3}, the valence band maximum (VBM) is at the $V$ point near $\Gamma$ on the $\Gamma$-$K$ path. The CBM is also on the $\Gamma$-$K$ path, but at a point $P$ near the midpoint of the path.

\begin{figure}
  \includegraphics[width=8cm]{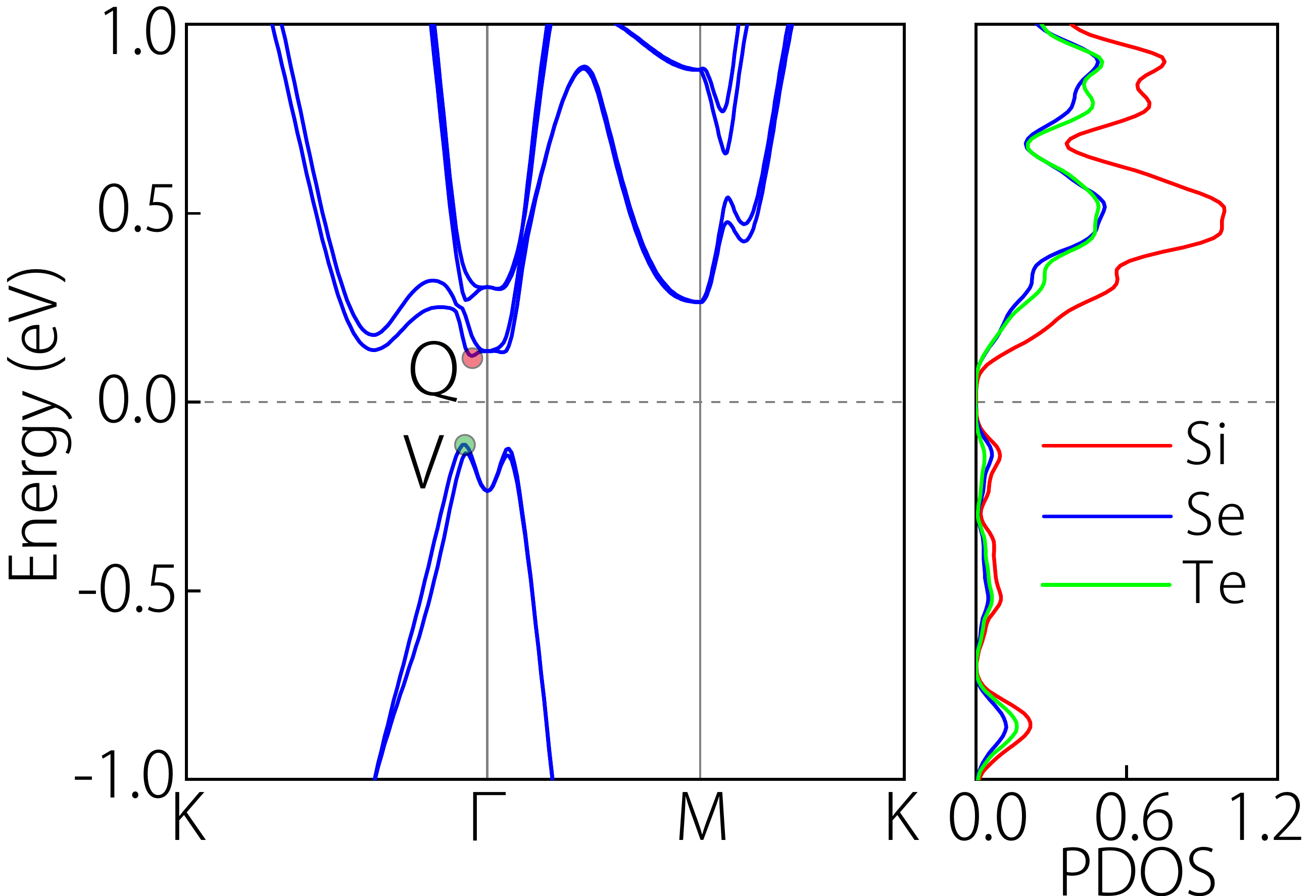}
\caption{Band structure and PDOS of monolayer $\mathrm{Si_2SeTe}$ with SOC included. Zero energy is set at the mid point of the band gap.}\label{fig4}
\end{figure}

The band structure with SOC included is shown in Fig.~\ref{fig4}. One can see that the system remains a semiconductor. The band gap is slightly reduced to about 235 meV. Interestingly, while the VBM position is more or less unchanged, the CBM changes from $P$ to a point $Q$ that is more close to $\Gamma$. In fact, here, the CBM and VBM are quite close to each other in Brillouin zone (BZ), so the system is almost a direct gap semiconductor.

Previous calculations on monolayer Si$_2$Te$_2$ predicted it to be a quantum spin Hall insulator~\cite{y9,y10}. Here, we have checked the $\mathbb{Z}_2$ invariant of monolayer Si$_2$SeTe~\cite{bc} (and also for the other two) by the Wilson loop method~\cite{w1,wcc} and found that it is trivial. Nevertheless, one observes that
SOC does bring notable changes to the band structure. Especially for the conduction bands around the $\Gamma$ point, there appears a large band splitting. The original doublet at $\Gamma$ in Fig.~\ref{fig3} now split into two doublets in Fig.~\ref{fig4} with a sizable shift in energy $\sim 170$ meV. This splitting is closely connected to the structural asymmetry in 2D JSD. In comparison,
in monolayer Si$_2$Te$_2$, the bands are spin degenerate due to the preserved inversion and time reversal symmetries.

\begin{figure}
  \includegraphics[width=8cm]{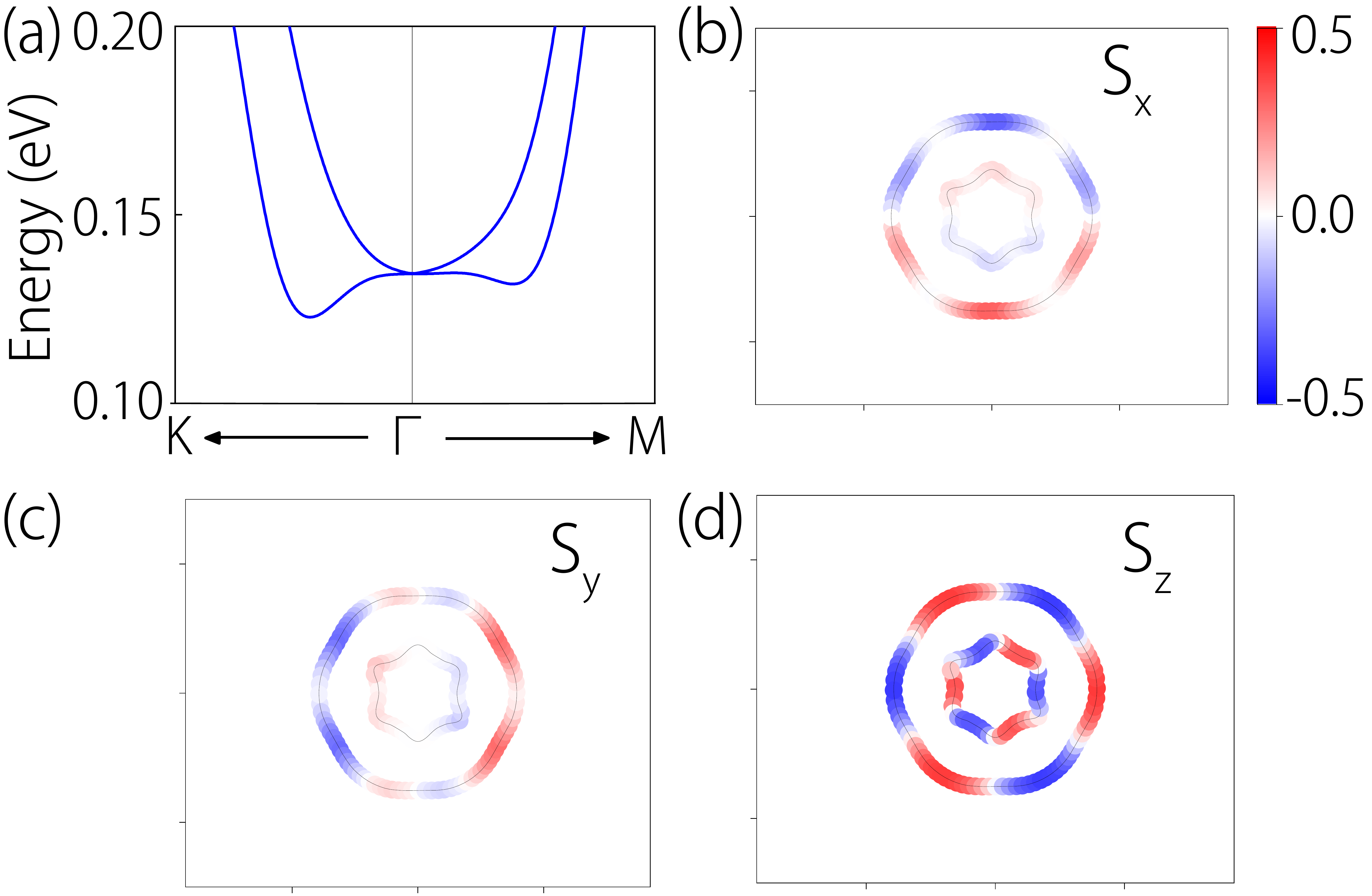}
\caption{(a) Enlarged view of the conduction band in Fig.~\ref{fig4} around the $\Gamma$ point. (b-d) show the spin polarization of states on the equi-energy surface at 0.14 eV. }\label{fig5}
\end{figure}

Let's focus on the lower doublet, which is close to the CBM. An enlarged figure about this doublet is shown in Fig.~\ref{fig5}.  The two bands form a Kramers degeneracy point at $\Gamma$ and their dispersion look similar to the Rashba model. In Fig.~\ref{fig5}(b-d), we plot the spin polarization on an equi-energy surface at 0.14 eV, close to the degeneracy point. Interestingly, besides in-plane spin polarization $S_x$ and $S_y$, we also find a sizable spin polarization $S_z$ in the out-of-plane direction. This is clearly beyond the simple linear Rashba model $H_R=\alpha (\bm k\times \hat{z})\cdot\bm\sigma$~\cite{y1}, which only has in-plane components. Here, $\alpha$ is the coupling strength, $\bm k$ the momentum is in the 2D basal plane, and $\bm\sigma$ is a vector of Pauli matrices corresponding to electron spin. To capture the emergence of $S_z$ component, we construct an effective $k\cdot p$ model expanded at $\Gamma$ by including higher order terms in the expansion. The doublet at $\Gamma$ corresponds to the $\Gamma_{4}$ double-valued representation of $C_{3v}$. Using the two states as basis, the generators for $C_{3v}$ can be represented as
\begin{equation}
  C_3=e^{-i\pi\sigma_z/3},\qquad M_y=-i\sigma_y,
\end{equation}
and the time reversal operator $\mathcal{T}=-i\sigma_{y}\mathcal{K}$ with $\mathcal{K}$ the complex conjugation.
These symmetries constrain the form of the effective Hamiltonian by
\begin{align}
  C_{3}H(\bm k)C_{3}^{-1}&=H(R_{3}\bm k),\\
  M_{y}H(\bm k)M_{y}^{-1}&= H(k_{x}, -k_{y}),\\
  \mathcal{T}H(\bm k)\mathcal{T}^{-1} &= H(-\bm k),
\end{align}
where the momentum $\bm k$ is measured from $\Gamma$.
Expanding $H(\bm k)$ to the $k^4$ order under the above constraints, we obtain the following $k\cdot p$ effective model:
\begin{align}
H_\text{eff}=&c_{1}(k_{x}\sigma_{y}-k_y\sigma_x)+c_{2}k^{2}\sigma_{0}+c_{3}(k_{x}\sigma_{y}-k_{y}\sigma_{x})k^{2}\nonumber\\
&+c_{4}(k_{y}^{3}-3k_{y}k_{x}^{2})\sigma_{z}+c_{5}k^{4}\sigma_{0},
\end{align}
where $\sigma_0$ is the $2\times 2$ identity matrix and $c$'s are the model parameters. One observes that the first term is just the linear Rashba SOC. Up to $k^2$ order, there is no contribution to the $S_z$ polarization. The $S_z$ polarization term only appears at the $k$ cubic order, in the fourth term here. The cubic terms are also related to the hexagonal warping effect.  The even order terms does not give a spin splitting, which is a result of the $\mathcal{T}$ symmetry. Using this model to fit the DFT band structure of monolayer Si$_2$Te$_2$, we obtain the following parameter values: $c_1=0.022$ eV$\cdot$\AA, $c_2=1.731$ eV$\cdot$$\mathrm{\AA}^2$, $c_3=25.262$ eV$\cdot$$\mathrm{\AA}^3$, $c_4=89.806$ eV$\cdot$$\mathrm{\AA}^4$ and $c_5=3.721$ eV$\cdot$$\mathrm{\AA}^5$.

\begin{table}
\centering \caption{Calculated piezoelectric coefficients for monolayer JSD materials. Here,  $e_{ij}$ is in unit of $10^{-10}$ C/m, and  $d_{ij}$ is in unit of pm/V. }\label{tab1}
  \begin{tabular*}{0.48\textwidth}{@{\extracolsep{\fill}}lcccc}
  \hline\hline
 Name &$e_{11}$&$e_{31}$&$d_{11}$&$d_{31}$\\\hline
$\mathrm{Si_2SSe}$ &16.58  &0.197   &27.73  &0.157                           \\
$\mathrm{Si_2STe}$  &30.30  &0.374   &62.36  &0.310                       \\
$\mathrm{Si_2SeTe}$ &16.91   &0.200   &40.52   &0.182
\\\hline\hline
\end{tabular*}
\end{table}

\begin{figure}
  \includegraphics[width=8cm]{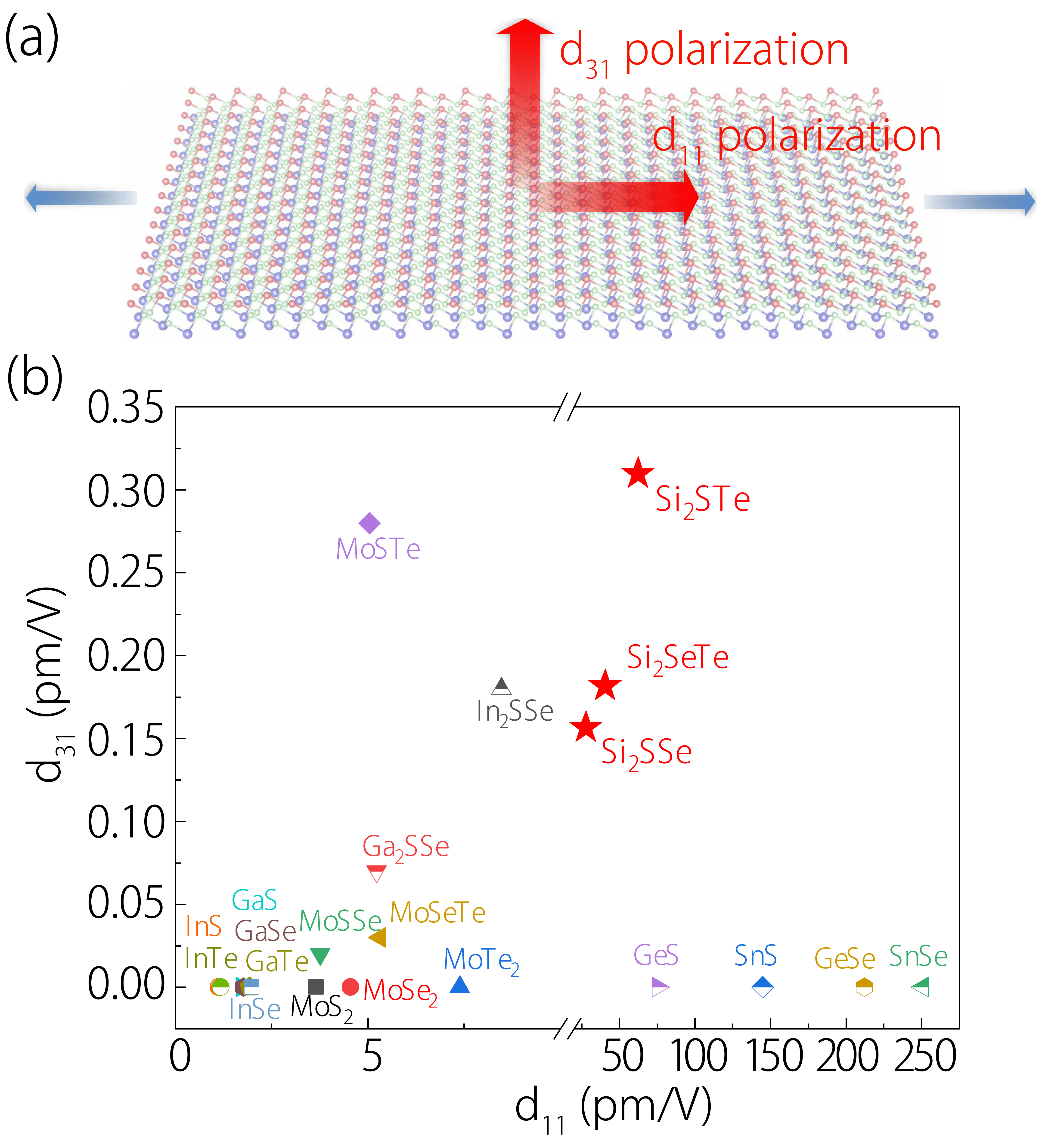}
  \caption{(a) Schematic figure showing the piezoelectric response in monolayer Si$_2$SeTe. Under a strain/stress along $x$, electric polarization is induced along $x$ and $z$, corresponding to $d_{11}$ and $d_{31}$, respectively. (b) Comparison of piezoelectric parameter values among well-known 2D piezoelectric materials~\cite{q4-2}. }\label{fig6}
\end{figure}

\section{Piezoelectricity}
Piezoelectricity results from the linear electromechanical interaction between the mechanical and electrical states.
It requires $\mathcal{P}$ symmetry breaking. Hence, it does not appear in monolayer Si$_2$Te$_2$ but naturally exists in monolayer JSDs.
The piezoelectric response can be described by the piezoelectric stress tensor $e_{ijk}$ or the piezoelectric strain tensor $d_{ijk}$, which are third rank tensors with $i,j,k=1,2,3$.  They are defined by the relations
 \begin{equation}\label{pe0}
      e_{ijk}=\frac{\partial P_i}{\partial \varepsilon_{jk}},\qquad
   d_{ijk}=\frac{\partial P_i}{\partial \sigma_{jk}},
 \end{equation}
in which $P_i$ is the electric polarization vector, $\varepsilon_{jk}$ and $\sigma_{jk}$ are the strain and stress tensors, respectively. Typically, one combines the latter two indices in $e_{ijk}$ and $d_{ijk}$ by using the Voigt notation and simplify them as $e_{i\ell}$ and $d_{i\ell}$, where $\ell=1,\cdots,6$. In practice, one first calculates $e_{i\ell}$, then $d_{i\ell}$ can be derived from the relation
\begin{equation}\label{ed}
  e_{i\ell}=\sum_{k}d_{ik}C_{k\ell}.
\end{equation}

For 2D systems, as we have mentioned, one usually only considers strain and stress in the basal plane. This eliminates the
elements in $e_{i\ell}$ and $d_{i\ell}$ for $\ell=3,4,5$. (The out-of-plane electric polarization, i.e., $i=3$, still needs to be considered.) For monolayer JSDs, the $C_{3v}$ symmetry requires the piezoelectric tensors taking the following form~\cite{q5-11,q5}
 \begin{equation}\label{pe1-1}
 e=\left(
    \begin{array}{ccc}
      e_{11} & -e_{11} & \cdot \\
       \cdot & \cdot & -e_{11} \\
      e_{31} & e_{31} & \cdot \\
    \end{array}
  \right),
\end{equation}
  \begin{equation}\label{pe1-2}
  d= \left(
    \begin{array}{ccc}
      d_{11} & -d_{11} & \cdot \\
      \cdot & \cdot & -2d_{11} \\
      d_{31} & d_{31} & \cdot \\
    \end{array}
  \right).
\end{equation}
Here, for simplicity, we dropped the three columns corresponding to $\ell=3,4,5$. The relation in (\ref{ed}) now takes the form of
\begin{equation}\label{pe2}
    d_{11}=\frac{e_{11}}{C_{11}-C_{12}},\qquad d_{31}=\frac{e_{31}}{C_{11}+C_{12}}.
\end{equation}

Here, for monolayer JSDs, the piezoelectric tensor $e_{i\ell}$ or $d_{i\ell}$ has only two independent element. $d_{11}$ ($e_{11}$) represents an in-plane response, i.e., a piezoelectric polarization along $x$ is induced by a stress (strain) along $x$ (see Fig.~\ref{fig6}(a)). The more interesting part is the out-of-plane response embodied by $d_{31}$ ($e_{31}$), which means the stress (strain) along $x$ can also produce a polarization in $z$ direction (see Fig.~\ref{fig6}(a)). Such a response is rare in 2D materials and was first proposed in Janus monolayer TMD materials~\cite{q5-11}. It was suggested that this out-of-plane piezoelectricity could promote the flexibility of piezoelectric device operations and the compatibility with the existing microelectronic device configurations. In Janus monolayer TMDs, $d_{31}$ is found to be on the order of 0.01 pm/V and $d_{11}$ is of a few pm/V~\cite{q5-11}.

Our results for monolayer JSDs are listed in Table~II. One observes that due to the stronger intrinsic field, the piezoelectric coefficients are also much larger than those of Janus TMDs. The values are overall an order of magnitude larger than Janus TMDs. Among the three JSDs, the strongest response is achieved in monolayer Si$_2$STe, with $d_{31}= 0.31 $ pm/V and $d_{11}=62.36$ pm/V. We note that this value of $d_{11}$ is larger than most reported 2D materials~\cite{q4-2}, perhaps only lower than some members of the SnSe family (such as 2D SnSe and GeSe). The magnitude of $d_{31}$ is also large among the existing 2D materials (although some predicted structures may achieve even higher values~\cite{g5,g7}). In Fig.~\ref{fig6}(b), we compare the piezoelectric properties among well-known piezoelectric 2D materials. One observes that the monolayer JSD materials stand out as the one that has large value in both $d_{11}$ and $d_{31}$. This indicates their great potential in device applications.

\section{Strain-induced transition}

2D materials generally can sustain much larger strains than 3D materials. Here, from the calculation of the strain-stress curve~\cite{bc}, we find that the monolayer JSDs have a linear elastic region extended to $\sim 8\%$ strain, and the critical strains can be $>20\%$. This again demonstrates the flexibility of these materials.

\begin{figure}
  \includegraphics[width=8cm]{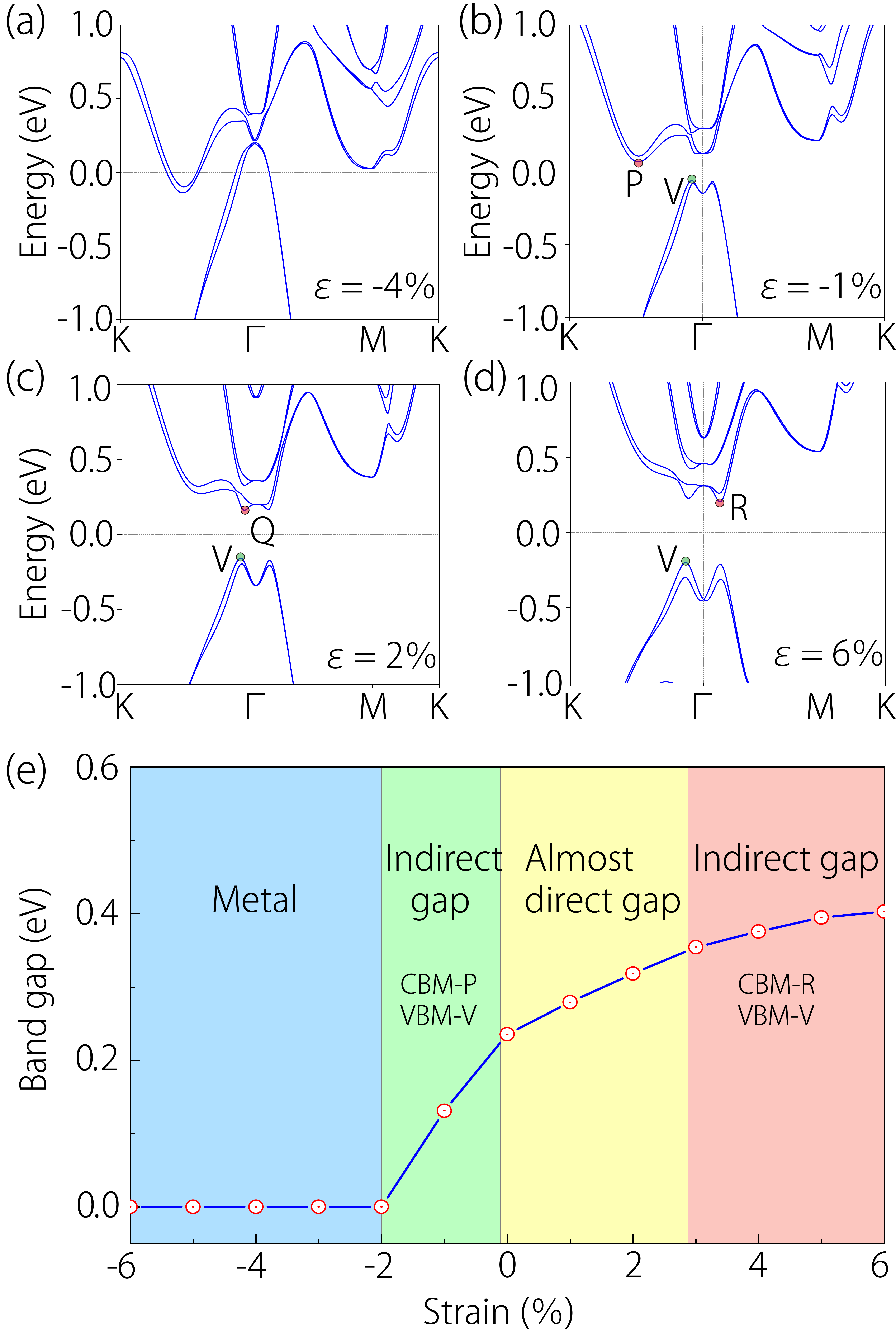}\caption{(a-d) Evolution of band structure of monolayer Si$_2$SeTe under different biaxial strains. (e) shows the variation of band gap under applied strain and a schematic phase diagram for monolayer Si$_2$SeTe.  }\label{fig7}
\end{figure}

Strain engineering has been widely used as an effective way to tailor the properties of 2D materials~\cite{h1,h2,h3,h4}. Here, we investigate the strain effects on the band structure. In Fig.~\ref{fig3} and \ref{fig4}, we have noted that for monolayer Si$_2$SeTe, there are two parts of the conduction band, $P$ and $Q$, which compete for the CBM. One naturally expects that a small strain can switch their order in energy and determine which one becomes the CBM. This is confirmed by our results in Fig.~\ref{fig7}, which shows the band structures under different biaxial strains. One observes that a small compressive strain $-1\%$ can already switch the CBM from $Q$ to $P$. Then the band gap types changes from almost direct to indirect. The band gap size also shrinks with compressive strain and even closes at $-2\%$ strain. On the tensile strain side, the $P$ valley moves up in energy, and the band gap increases with strain. Interestingly, at about 3\% strain, the CBM changes from $Q$ to another valley $R$ on the $\Gamma$-$M$ path (see Fig.~\ref{fig7}(d)). These results are summarized in Fig.~\ref{fig7}(e).

\begin{figure}
  \includegraphics[width=8cm]{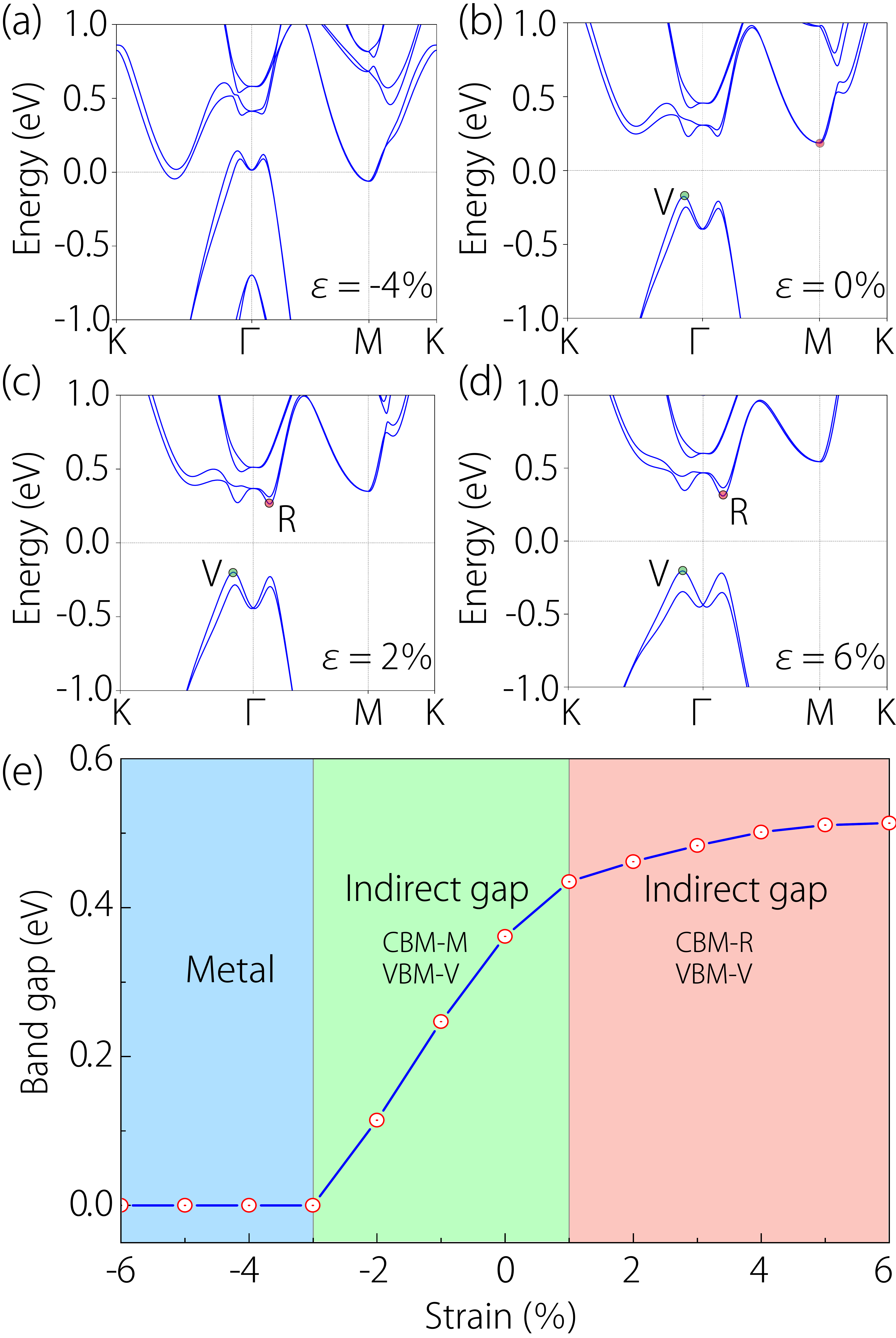}\caption{(a-d) Evolution of band structure of monolayer Si$_2$STe under different biaxial strains. (e) shows the variation of band gap under applied strain and a schematic phase diagram for monolayer Si$_2$STe.  }\label{fig8}
\end{figure}

The case in monolayer Si$_2$STe is different from Si$_2$SeTe. As shown in Fig.~\ref{fig8}(b), without strain, monolayer Si$_2$STe is a semiconductor with an indirect band gap $\sim 361$ meV. Its VBM is at a similar $V$ point like monolayer Si$_2$SeTe. However, its CBM is at $M$ point. Under compressive strains, the band gap decreases and vanishes at about $-3\%$ strain. As for tensile strains, there is a transition at about $1\%$ strain where the CBM switches from $M$ to $R$, as shown in Fig.~\ref{fig8}(c,d). A schematic phase diagram is shown in Fig.~\ref{fig8}(e).

Such transitions in the CBM location could result in notable physical effects. For example, the accompanied change in the band gap type makes a big difference in the optical properties, especially the luminescence properties of a semiconductor.
In addition, the different valleys have different effective mass, which directly affects the carrier transport. For instance, for monolayer Si$_2$SeTe, the shift of CBM from $P$ valley to $Q$ valley decreases the electron effective mass by almost five times from about $0.7m_e$ to about $0.15m_e$ where $m_e$ is the bare electron mass. This change will manifest as a jump in the conductivity under strain.

\section{Discussion and Conclusion}

In this work, we have revealed many interesting properties of 2D JSDs. Derived from the Janus structure, the strong spin splitting and the large piezoelectric response make these materials promising for device applications. Besides conventional concepts such as spin transistors/modulators, mechanical sensors, actuators, power transducers, etc., combining the two characters may generate some novel device concepts like some piezoelectric controlled spin devices.

Experimentally, monolayer Si$_2$Te$_2$ has been successfully grown on Sb$_2$Te$_3$ substrate~\cite{y8}. The monolayer JSDs Si$_2$STe and Si$_2$SeTe may be synthesized based on Si$_2$Te$_2$, following the similar approach as for monolayer Janus TMDs.
For example, they may be obtained via sulfurization or selenization of Si$_2$Te$_2$ on substrate at elevated temperatures, due to the stronger electronegativity of S/Se than Te. Another way is to first strip off the top Te layer and replace them with H using a remote H plasma, and then proceed with thermal sulfurization or selenization. These methods were successfully demonstrated in previous works on synthesizing Janus TMDs~\cite{e1,e2}. As for strain application, in practice, the compressive strain can be applied by choosing a proper substrate with a matching lattice but smaller lattice constant. Applying tensile strain on 2D materials is more convenient. It can be done by a beam bending apparatus or by using a stretchable substrate~\cite{h5,h6}.

In conclusion, we have proposed a family of Si based 2D materials, the 2D JSD materials, which could be realized from monolayer Si$_2$Te$_2$ using similar techniques for the synthesis of Janus TMDs. We demonstrate the good stability of these materials. We show that the their intrinsic polar field can be even stronger than Janus TMDs. Due to this strong inversion symmetry breaking, they manifest sizable SOC splitting and pronounced piezoelectric response. The SOC splitting acquires a large out-of-plane component which is beyond the linear Rashba model. The piezoelectric tensor has large value in both $d_{11}$ and $d_{31}$, making these materials distinct among the existing 2D materials. We work reveals a new 2D family of materials that are compatible with Si-based technology and have great potential for spintronic and piezoelectric device applications.

\begin{acknowledgments}
The authors thank D. L. Deng for helpful discussions. This work is supported by Singapore NRF CRP22-2019-0061 and Natural Science Basis Research Plan in Shaanxi Province of China (2021JM-456). We acknowledge support from Shanxi Supercomputing Center of China, and some calculations were performed on TianHe-2.
\end{acknowledgments}

\end{document}